\documentclass[letter,twocolumn,showpacs,aps]{revtex4}

\expandafter\let\csname equation*\endcsname\relax
\expandafter\let\csname endequation*\endcsname\relax

\usepackage{amsmath}
\usepackage{amssymb}
\usepackage{color}
\usepackage{verbatim}
\usepackage{array}
\usepackage{graphicx}
\usepackage{epsfig}
\usepackage{bm}
\usepackage[ragged]{sidecap}
\usepackage{dsfont}
\usepackage{ulem}
\usepackage{mathrsfs}
\usepackage{setspace}

\usepackage{subfigure}

\newcommand{\ds}{\displaystyle}

\newcommand{\Red}[1]{{\color{black}{ #1}}}

\renewcommand{\vec}{\bm}

\newcommand{\opvector}[1]{\hat{\mathbf{#1}}}
\newcommand{\opscalar}[1]{\hat{\textrm #1}}

\newcommand{\paulix}{\left(\begin{array}{cc}0&1\\1&0\end{array}\right)}
\newcommand{\pauliy}{\left(\begin{array}{cc}0&-i\\i&0\end{array}\right)}

\begin{document}
\preprint{\hfill\parbox[b]{0.3\hsize}{ }}

\title{Constructing a nanoscale optical polarizer with a graphene stack
}

\author{F. Fratini\footnote{
E-mail addresses: fratini.filippo@gmail.com
}}
\affiliation
{\it
University of Applied Sciences BFI Vienna, Wohlmutstra\ss e 22, A-1020 Wien, Austria
}

\date{\today}

\begin{abstract}
Two-dimensional graphene layers exhibit many fascinating properties which have sparkled into applied research with the aim to build innovative electronic devices. Here, we theoretically demonstrate that, when the carriers velocity is constrained along one direction, a monoatomic graphene layer exhibits dichroism. A fraction $2\pi\alpha$ of the light polarized along the carriers motion is absorbed, while light polarized perpendicularly to that direction is not absorbed. A stack of two-dimensional graphene layers whose carriers velocity is constrained along one direction (such as a stack subjected to a suitable gate voltage) is thus able to polarize light up to the wished degree by selective absorption. These findings pave the way for built-in controllable optical polarizers in upcoming graphene nanoscale devices.
\end{abstract}


\maketitle

%

The rise of graphene in recent years is due to its exciting, unusual electronic properties \cite{Cas2009}. It is a zero-gap semiconductor that is as thin as one atom and that was supposed not to exist \cite{Gei2007}. Although only one-atom thick, it shows remarkably high electron mobility and minimal conductivity, as well as other unusual quantum electrodynamical properties \cite{Nov2004, Nov2005, Nov2006, Zha2005, Kat2006a, Kat2006b}. Harnessing these features into technology is the tantalizing goal of nowadays applied research \cite{Bon2010}.

\medskip 

Modern fast communication relies on optical transport and microscopic devices. This draws interest in flexible devices which are capable of light transport or light control at the nanoscale \cite{Fratini2014b, Fratini2016}. In this flourishing research field, Graphene is seen as one of the ideal candidates. The optical transparency of a Graphene monoatomic layer has been shown to be uniquely determined by the fine structure constant and not by material parameters \cite{Nai2008}. Such an optical transparency has been found to hold for visible light of any frequency and at room temperature \cite{She2009}. This unique and flexible optical characteristics, jointly with the microscopic dimensions of graphene layers ($\sim$ 0.35 nm thick), open up the possibility of using graphene in upcoming nanoscale optical devices to be used in optoelectronics and quantum processing.
Toward that goal, in this Letter we show that a monoatomic graphene layer manifests dichroic properties when the carriers velocity is constrained along one direction. 
A fraction $2\pi\alpha$ of the light polarized along the carriers direction of motion is absorbed, while the layer is completely transparent to light polarized perpendicularly to the carriers direction of motion. 
Thus, the polarization selective absorption of a graphene stack is proportional to the number of layers, each layer absorbing $2\pi\alpha\approx4.6$\% over the visible spectrum. The regime from unpolarized to fully polarized light is controlled by piling up from $0$ to $\sim 100$ graphene layers, as we shall show.

\medskip

We start by recalling that the low-energy excitations in graphene can be efficiently described by the Dirac Hamiltonian for massless particles \cite{Cas2009}:
\begin{equation}
\label{eq:Hmin}
\opscalar H_G=v_F\,\opvector q\cdot \vec \sigma~,
\end{equation}
where $v_F$ is the Fermi velocity, $\opvector q$ is the crystalline momentum operator and $\opscalar h=\frac{\opvector q\cdot \vec \sigma}{|\vec q|}$ is the pseudo-helicity operator that selects the band to which the electron belongs. $\opscalar h$ has eigenvalues $h=-1$ and $+1$ for electrons in the valence and conduction band, respectively. The spectrum is found by solving $\opscalar H\Psi(\vec r)=E\Psi(\vec r)$. Solutions that have defined pseudo-helicity can be written as
\begin{equation}
\label{eq:WF}
\begin{array}{lcl}
\Psi^{\vec q, h}(\vec r, t)&=&\ds\xi_{h}^{\vec q}\;\frac{e^{+\frac{i}{\hbar}(\vec q\vec r-Et)}}{\sqrt{V}}~,
\end{array}
\end{equation}
where $\bm q=(q_x, q_y, q_z)$ is the electron crystalline momentum, $\xi_{h}^{\vec q}$ defines the SU(2) pseudo-helicity state, $V$ is the voume where the wavefunction is defined, while $E=\pm v_F\,|\vec q|$ is the energy for electrons belonging to valence ($-$) and conduction ($+$) band. The pseudo-helicity state $\xi^{\vec q}_h$ is defined by giving the momentum direction ($\vec q/|\vec q|$) and the band ($h$). For instance, an electron that has momentum along the $x$-axis and that belongs to the valence band ($h=-1$), has its pseudo-helicity state $\xi_{-}^{\vec q}=\frac{1}{\sqrt{2}}\left(\begin{array}{c}1\\-1\end{array}\right)$. More generally, by using the Euler rotation theorem \cite{Sak1994}, we may express the pseudo-helicity state for an electron belonging to the valence ($h=-1$) or conduction ($h=+1$) band, and possessing a momentum direction $\vec q/|\vec q|=(\sin\theta\cos\varphi,\sin\theta\sin\varphi,\cos\theta)$ as 
\begin{equation}
\xi_{-}^{\vec q}=\left(
\begin{array}{c}
e^{\frac{i\varphi}{2}}\cos(\theta/2)\\
-e^{-\frac{i\varphi}{2}}\sin(\theta/2)
\end{array}
\right)\;,\;
\xi_{+}^{\vec q}=\left(
\begin{array}{c}
e^{\frac{i\varphi}{2}}\sin(\theta/2)\\
e^{-\frac{i\varphi}{2}}\cos(\theta/2)
\end{array}
\right)~.
\end{equation}
Since electrons are constrained in the two-dimensional graphene layer (which is here taken as the $xy$-plane), only two values are allowed for $\theta$: $\theta=\pm\pi/2$.

\begin{figure}[b]
\includegraphics[scale=0.3]{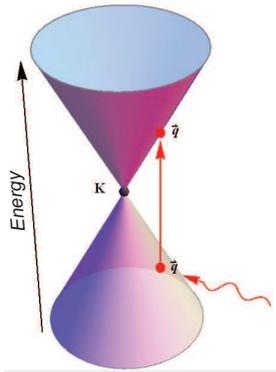}
\caption{(color online). Dirac-cone band structure of low-energy excitations in graphene around the Dirac point $\bf K$. As a consequence of a photon absorption, the electron is promoted from the valence band (lower cone) to the conduction band (upper cone). The crystalline momentum $\vec q$ of the electron is conserved during the transition.
}
\label{fig:Cone}
\end{figure}

\medskip

The Fermi Golden Rule (FGR) has been proved to successfully describe graphene optical properties within the efficient model in Eq. \eqref{eq:Hmin} \cite{Nai2008}. Following such a success, in this Letter we shall make large use of it for deriving polarization selective absorption properties of graphene layers. The rate for photon absorption by a single graphene layer within the FGR can be expressed as \cite{Nai2008, Fratini2011}:
\begin{equation}
\label{eq:TR}
\Gamma=\ds\frac{4\pi^2v_F^2\,I\,\alpha}{\omega^2}\;
\Big|\mathcal{M}\,\Big|^2\,\delta(E_f-E_i-E_\gamma)
\end{equation}
where $\omega$ and $I$ are the angular frequency and the intensity of incident light, $\alpha$ is the fine structure constant, while $\mathcal{M}$ is the amplitude for the process.
Pauli blocking forbids transitions within the valence band. Therefore, if any transition is to be generated by photon absorption, it must be from the valence to the conduction band. These transitions are called `vertical transition' (see Fig. \ref{fig:Cone}). 
Moreover, the graphene minimal cell has the shape of hexagon with a size of the order of $\approx 2$ \AA. Since the electron is confined in that hexagon, and given that visible light has wavelength from $400$ to $700$ nm, we may safely apply dipole approximation.
With these considerations, the squared amplitude for that process can be written as
\begin{equation}
\label{eq:M}
\begin{array}{lcl}
\Big|\mathcal{M}\,\Big|^2&=&\ds\Big|(\xi_{+}^{\bm q_f})^\dag \, \bm\sigma\cdot\bm\epsilon \; \xi_{-}^{\bm q_i}\Big|^2\quad
\frac{(2\pi\hbar)^3}{V}\,\delta^3(\bm q_f-\bm q_i) 
\end{array}
\end{equation}
where $\bm \epsilon$ is the polarization unit vector of the incident light \cite{Fratini2011}. From \eqref{eq:M} we see that the Dirac delta function forces the final electronic crystalline momentum to be equal to the initial one. Our model thus correctly reflects the fact that optically generated transitions in graphene near the Dirac points are vertical, i.e. they proceed from a state in the valence band to a state in the conduction band, where the two states possess the same crystalline momentum (Fig. \ref{fig:Cone}). 

\begin{figure}[t]
\includegraphics[scale=0.4]{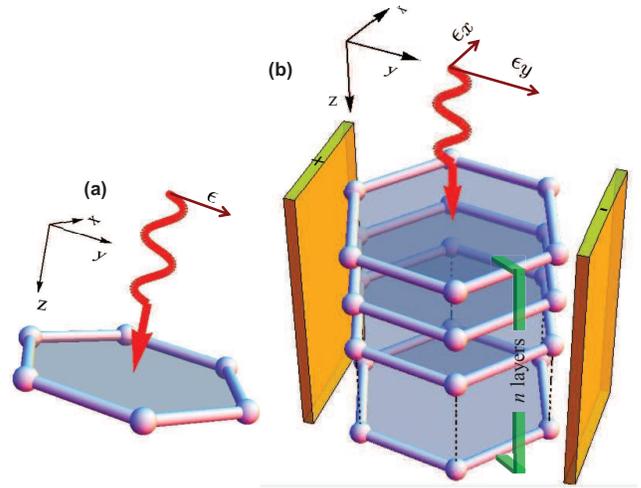}
\caption{(color online). (a) Linearly polarized light that impinges orthogonally onto the graphene layer; (b) Unpolarized light that impinges orthogonally onto a graphene stack subjected to a gate voltage. A fraction $2\pi\alpha$ of the polarization component orthogonal to the electric field direction is absorbed at each layer. Upon crossing $\sim 100$ layers, light is fully linearly polarized along the direction of the electric field (see also Fig. \ref{fig:DLP}).
}
\label{fig:StackSingle}
\end{figure}

\medskip 

It is well known that the same matrix amplitude \eqref{eq:M} applies also to stimulated emission. This can be derived either by direct calculation or by applying the principle of detailed balancing \cite{Bransden}. 
The rate of stimulated emission is equal to that one in Eq. \eqref{eq:TR} times a weighting factor accounting for the fraction of population inversion that is present when the photon hits the graphene layer. 

\medskip 

Let us come back to Eq. \eqref{eq:M}. We notice that the matrix element $T\equiv\Big|(\xi_{+}^{\bm q_f})^\dag \, \bm\sigma\cdot\bm\epsilon \; \xi_{-}^{\bm q_i}\Big|$ strongly depends on the direction of the electronic crystalline momentum with respect to the light polarization. 
In other words, optical absorptions are inhomogeneous around the Dirac points. This fact has been for the first time pointed out by Gr\"uneis et al. already in 2003 \cite{Gru2003}.
To make a clear example, let us suppose that the crystalline momentum of graphene electrons is constrained along the $x$-axis. Let us further suppose that the incident light has direction along the $z$-axis, and linear polarization along the $y$-axis (i.e., light is incident orthogonally onto the graphene layer and is polarized orthogonally to the electron direction). The situation is depicted in Fig. \ref{fig:StackSingle}, panel (a). For this situation, we get
$T=\Big|\frac{1}{2}\big(1\;1\big) 
\, \pauliy \, 
\left(\begin{array}{c}1\\-1\end{array}\right)\Big|=1$. Substituting this result in Eqs. \eqref{eq:TR} and \eqref{eq:M}, multiplying by the density of final electron states as given by Peres \cite{Peres}, we get the rate of photon absorption $\Gamma=2A_c\pi\alpha I/(\hbar \omega) $, where $A_c$ is the area of the hexagon cell as defined in \cite{Peres}. The absorbed light intensity, being the rate of absorbed energy per unit cell $A_c$, can be written as $I_{abs}=\hbar\omega\Gamma/A_c=2\pi\alpha I$, which yealds the ration $I_{abs}/I=2\pi\alpha$.
On the other hand, if we suppose that the incident light has polarization parallel to the electron direction (i.e., along the $x$-axis), we get
$T=\Big|\frac{1}{2}\big(1\;1\big) 
\, \paulix \, 
\left(\begin{array}{c}1\\-1\end{array}\right)\Big|=0$. This leads to $I_{abs}/I=0$.

\medskip 

From the foregoing discussion, we conclude that only light whose polarization is orthogonal to the electrons crystalline momentum is partially absorbed. 
In order to explore the dichroic properties of graphene, we must therefore constrain the electron cristalline momentum along a given direction. This can be done, for example, by the application of a gate voltage of a few tens of mV \cite{Feng2008, Xia2010, Li2008, Yu2012, Ch2012}, as we shall consider below. Such a small voltage does not significantly change the band structure of graphene \cite{Xia2010}, yet it is strong enough to constrain the electron velocity along the gate electric field direction (electric current has been in fact measured with such a potential \cite{Peres}). 

\medskip 

Without any constraint for the crystalline momentum of electrons (or with unpolarized light), the absorption rate must be averaged over the crystalline momenta (or over the light polarizations) \cite{FratiniSc2011}. This would yield a half of the result above: $I_{abs}/I=\pi\alpha$. Such absorption rate has been in fact confirmed experimentally \cite{Nai2008, Bon2010}.

\medskip 

To best exemplify our findings, let us now suppose that unpolarized light with intensity I is incident on a stack of $n$ layers of graphene, where the stack is subjected to a suitable gate voltage so that the electrons crystalline momentum are constrained along one direction. For simplicity, this situation is depicted in Fig. \ref{fig:StackSingle}, panel (b). For the {\it incident} light, the intensity of the polarization components that are orthogonal ($\perp$) or parallel ($\parallel$) to the gate voltage is naturally the same: $I_\perp^0=I_\parallel^0=\frac{I}{2}$. Light will then cross $n$ graphene layers. Following the analysis we carried out above, the intensity of the polarization components of the {\it outgoing} light will be
\begin{equation}
\begin{array}{l}
\ds I_\perp^n=\big( 1 - 2\pi\alpha \big)^{n}I_\perp^0=\big( 1 - 2\pi\alpha \big)^{n}\,\frac{I}{2} ~,\\
\ds I_\parallel^n=I_\parallel^0=\frac{I}{2} ~.
\end{array}
\end{equation}

\begin{figure}[b]
\includegraphics[scale=0.6]{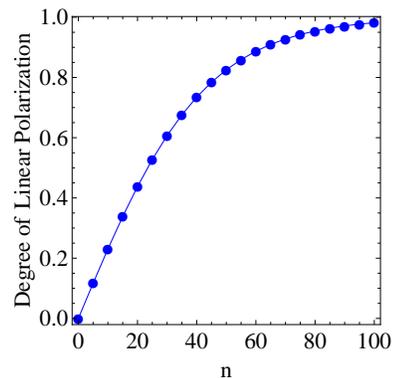}
\caption{(color online). Degree of linear polarization of light which has passed through a stack of $n$ layers of graphene on which a gate voltage of a few tens of mV is applied. The light incident onto the stack is unpolarized.
}
\label{fig:DLP}
\end{figure}

\medskip 

In Fig. \ref{fig:DLP}, the degree of linear polarization of outgoing light, $(I_\parallel^n-I_\perp^n)/(I_\parallel^n+I_\perp^n)$, is plotted against the number of layers of the stack, $n$.
The orthogonal polarization is extinguished after approx $\sim 100$ layers. In other words, upon the passage of light through a graphene stack of $\sim 100$ layers, only the polarization parallel to the electric field of the gate survives. The outgoing light is thus fully polarized.  

\medskip 

Although our analysis has been conducted for light absorption, a similar calculation applies for stimulated emission, as we already highlighted soon after Eq. \eqref{eq:M}. In this case, the curve in Fig. \ref{fig:DLP} will describe the degree of light amplification given by the Graphene layer to the light polarized along the applied electric field. We would have then a polarization dependent light amplification by stimulated emission of radiation. However, the rate of stimulated emission will need to be scaled by the fraction of population inversion that characterizes the Graphene stack. 
Exploring dichroism in the case of stimulated emission with the formalism developed here would be probably even easier than in the case of light absorption. That is because constraining the electronic crystalline momentum along a given direction in the conduction band is easier than in the valence band. 

\medskip 

The optical response of graphene subjected to a gate voltage has been studied in a number of works \cite{Feng2008, Xia2010, Li2008}. Also the dependence of the polarization absorption on the strength of an applied gate voltage \cite{Yu2012} or on the in-plane polarization axis \cite{Ch2012} has been studied. 
Here we have presented an analysis based on the Fermi Golden Rule that is able to capture graphene dichroism properties that were not reported previously.
Our results may be used in manifacturing built-in controllable polarizers to be installed in upcoming graphene based nanoscale devices. Such polarizers might be used for controlling the polarization of light at the nanoscale level, in upcoming photonic and quantum communication devices \cite{Bon2010, Fratini2014b}. For example, Graphene stacks could be used as a replacement of liquid crystals, which are currently used to change light polarization.

\end{document}